\documentclass[pre,aps,onecolumn]{revtex4-1}
\usepackage{amsmath,natbib,amsfonts}
\usepackage{amssymb,color}
\usepackage{graphicx}
\newcommand{\beqa}{\begin{eqnarray}}
\newcommand{\eeqa}{\end{eqnarray}}

\newcommand{\lan}{\langle}
\newcommand{\ran}{\rangle}

\begin{document}
\title{Extracting Work from a single heat bath using velocity dependent feedback.} 
\author{Arnab Saha$^1$, Rahul Marathe$^2$ and A. M. Jayannavar$^{3,4}$} 
\email{sahaarn@gmail.com, maratherahul@physics.iitd.ac.in, jayan@iopb.res.in}
\affiliation{$^1$Department of Physics, Savitribai Phule Pune University, Ganeshkhind, Pune  411007, India.\\
$^2$Department of Physics, Indian Institute of Technology, Delhi, Hauz Khas 110016, New Delhi, India.\\
$^3$Institute of Physics, Sachivalaya Marg, Bhubaneshwar 751005, Odhisha, India.\\
$^4$Homi Bhabha National Institute, Training School Complex, Anushakti Nagar, Mumbai 400085, India. }

\date{\today}
\begin{abstract}
{\textcolor{black}{Thermodynamics of nanoscale devices is an active area of research. Despite their noisy surrounding they often produce mechanical work (e.g. micro-heat engines) or display rectified Brownian motion (e.g. molecular motors). This invokes the research in terms of experimentally quantifiable thermodynamic efficiencies. To enhance the efficiency of such devices, close-loop control is an useful technique. Here a single Brownian particle is driven by a harmonic confinement with time-periodic contraction and expansion, together with a velocity feedback that acts on the particle only when the trap contracts. Due to this feedback we are able to extract thermodynamic work out of the system having single heat bath without violating the Second Law of Thermodynamics. We analyse the system using stochastic thermodynamics.}}
\end{abstract}
\maketitle

{\textcolor{black}{The Kelvin's statement of the Second Law of Thermodynamics reads ``it is impossible to extract heat from a single heat bath which can be used to perform useful work''\cite{Callen}. This can be avoided with a closed-loop control over the process by feedback mechanism. Feedback controls are applied to many systems to improve their performance. Experimentalists of natural sciences often use feedback to minimize vagaries of environmental perturbations. For example, in Atomic force microscopy thermal noise of the cantilever is reduced by a feedback like mechanism \cite{Liang00,Tamayo01}. For a systematic study on the behavior of any complex system, that depends on several externally tunable parameters, often one needs to vary one of the parameters, keeping others constant. It is a subject dealt with control theory, that heavily relies on feedback mechanisms at various levels. Biological systems and/or processes, being highly complex, often uses feed backed loops to enhance their efficiency and stability. Transport in cellular as well as in tissue scale via ion channels and motor proteins is one of the major examples \cite{Julicher97,Hawkes02,Qian02} where extensive usage of feed backed control has been perceived.}}

{\textcolor{black}{In the present paper we apply velocity dependent feedback control to a Brownian particle attached to a single heat bath and confined within a harmonic trap having time-periodic trapping strength to extract work from the system. Major examples of theoretical proposals to extract work from a single heat bath using information are Maxwell demon and Szilard engine. Recently these micro machines have also been realised experimentally \cite{jonne,mihai,koski}. The thermodynamics of work-extraction with a Brownian particle attached to a single heat bath, using information-aided feedback loop has witnessed recent significant developments. Nondeterministic Feedback usually depends on outcome of measurement of the state of the system and error occurred during the measurement process \cite{abreu1,priyo,kundu12}. In our current study, the velocity dependent feedback, which is instantaneous in nature, is applied only when the strength of the trap, an externally controllable parameter, increases. As the velocity-feedback employed here is independent of any measurement and related error, it is deterministic in nature \cite{Qian04,kim04}. As the feedback, we apply a drag force proportional to the instantaneous velocity of the particle and it acts opposite to the direction of the velocity. Therefore it is similar to the Stokes' drag acting on the particle but the crucial difference is the corresponding proportional coefficient is an external input parameter to the system which is not related to the temperature of the bath by fluctuation-dissipation relation. In other words, in presence of the feedback, the effective friction coefficient of the particle explicitly breaks the fluctuation-dissipation relation. Using the feedback we are able to extract work from the system as we can extract work using micro-heat engines with thermal bath (e.g.\cite{Edgar16,Arun14}) or with bacterial bath\cite{Sood16}. But we note that the trapped Brownian particle in \cite{Arun14,Edgar16} working as a micro-heat engine are time-periodically driven between two heat baths having different temperatures and the trap contracts (expands) when the particle is in contact with the cold (hot) bath. On the other hand, here, throughout all the cycles of contraction and expansion of the trap, the particle is always in contact with a single heat bath. Though the particle is slowed down (or, in analogy with micro heat-engines, it is {\it{cooled down}}) by the feedback only during the contraction of the trap and therefore it produces work.}}{\textcolor{black}{The Here we determine distributions $P(W)$ and $P(\eta)$ together with the averages of work $W$ and stochastic efficiency $\eta$, numerically calculated along the trajectories and for various cycles, using stochastic thermodynamics. The stochastic efficiency distribution is broad and almost unimodal (i.e. without any prominent local minima at finite efficiency). Its tail behaves as $\sim \eta^{-\alpha}$ with $\alpha\simeq 2$.}}

{\textcolor{black}{In the following sections first we will explain our model and implementation of the feedback in detail. In section (III) we will explain the numerical method to solve the model equations as well as the procedure to calculate stochastic thermodynamic quantities. In next section we analyse our results with physical interpretation and finally we conclude with discussion on our results.}} \\


{\textcolor{black}{We consider a single Brownian particle confined in a Harmonic trap. The trap strength is time-periodic and used as a protocol to drive the particle. The protocol used here is same as the protocol used in \cite{Arun14}. The equation of motion of the particle is given by the under-damped Langevin equation:
\beqa
m\ddot{x} = -\gamma \dot{x} -k(t)x+\sqrt{\gamma T} \xi(t),
\label{Lang1}
\eeqa
where, $m$ is the mass of the particle, $\gamma$ the friction coefficient, $T$ temperature of the bath, $k(t)$ is time-dependent trap strength. The noise $\xi(t)$ due to the heat bath is modeled as Gaussian white noise satisfying $\lan\xi(t)\ran=0$ and $\lan\xi(t)\xi(t')\ran=2\delta(t-t')$. In all the calculations we keep mass $m$ and the Boltzmann constant $k_B$ as unity, and all energies are measured in these units. The strength of the confinement $k(t)$ is varied with time in a cycle of duration $\tau$. In the first step $k(t)$ is decreased linearly from the initial value $k$ to $k/2$ as,
\[ k(t)=k\left(1-\frac{t}{\tau}\ \right)=k_1(t). ~~~~~~~0<t<\tau/2 \]
This step is the expansion step. After this the trap strength is decreased further to $k/4$ instantaneously, and since this step is instantaneous no heat is exchanged between the system and the bath. In the third step $k(t)$ is increased linearly from $k/4$ to $k/2$ as,
\[ k(t)=k\frac{t}{2\tau}\ =k_2(t), ~~~~~~~\tau/2<t<\tau \]
this is the compression step with temperature $T$. In this step the velocity feedback is employed by modifying the friction term in Eq. \ref{Lang1} to $\gamma_{fb}+\gamma$, where $\gamma_{fb} >0$ is the feedback parameter. Thus with this modified friction Langevin equation Eq. \ref{Lang1} during the compression step becomes:
\beqa
m\ddot{x} = -(\gamma_{fb}+\gamma) \dot{x} -k(t)x+\sqrt{\gamma T} \xi(t).
\label{Lang2}
\eeqa
For this equation Einstein's fluctuation dissipation relation namely $D\gamma=k_BT$ is not satisfied unlike in the case of Eq. \ref{Lang1} and this is the crucial step which allows us to extract heat from the heat bath which can be used to perform useful work. In the last step, the trap strength is finally increased to its initial value $k$ from $k/2$ instantaneously, without allowing any heat exchange with the bath. In the beginning of the last step the friction coefficient is restored to $\gamma$ so that the system is ready for the next expansion step. This cycle is then repeated.}}   
\begin{figure}[ht]
\centering
\includegraphics[width=1.0\columnwidth]{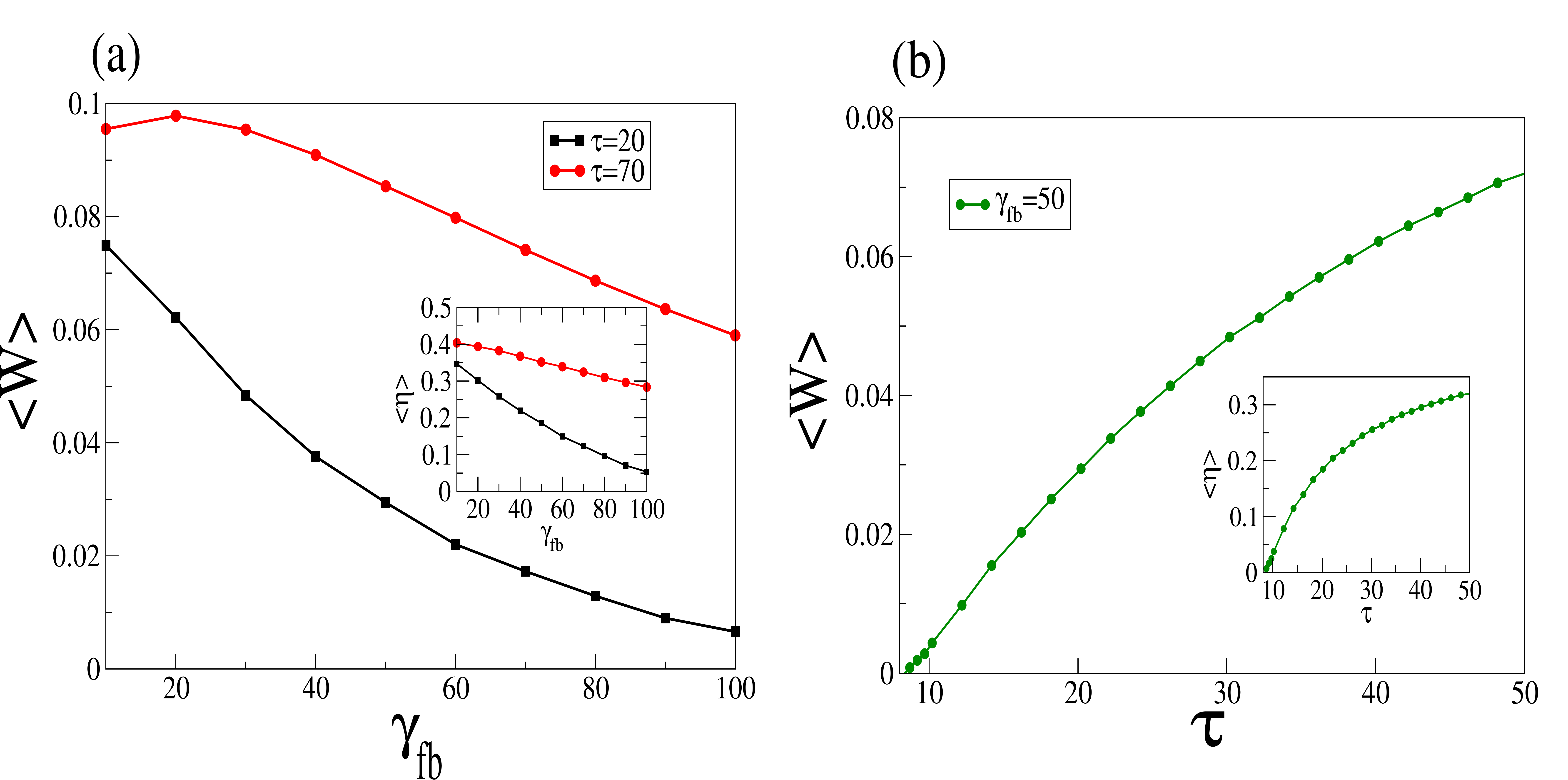}
\caption{(color online) (a) Plot of average work $\lan W\ran$ as a function of feedback parameter $\gamma_{fb}$ for two different cycle periods $\tau=70$ and $\tau=20$.The average work done is positive implying system does work on the external agent thus performing as an engine.  Inset efficiency $\lan\eta\ran$ as a function of $\gamma_{fb}$. (b) Plot of average work $\lan W\ran$ as a function of cycle period $\tau$. Inset efficiency $\lan\eta\ran $ as a function of $\tau$. Here $\gamma_{fb}=50$}
\label{fig1}
\end{figure}

\begin{figure}[ht]
\centering
\includegraphics[width=1.0\columnwidth]{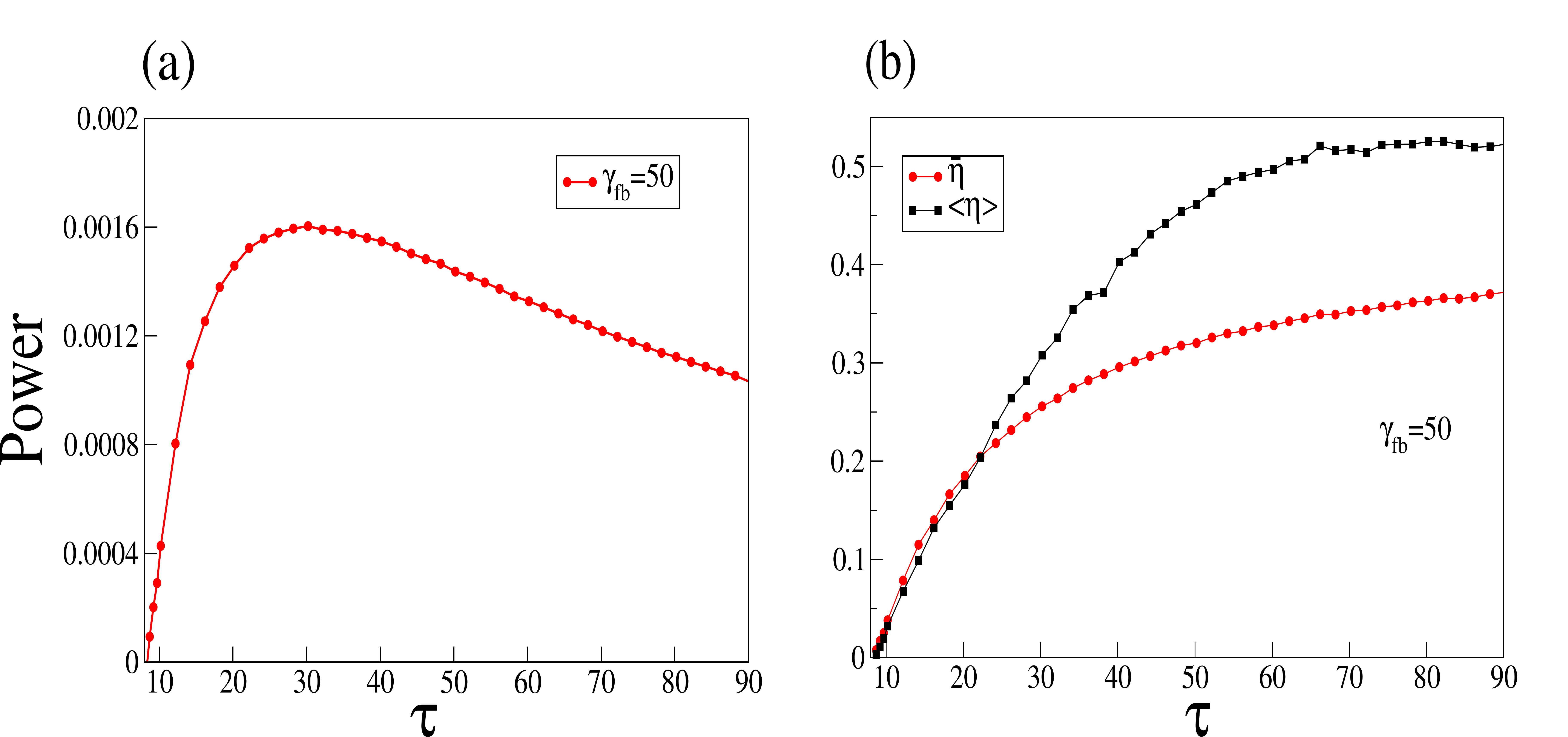}
\caption{(color online) (a) Plot of Power $\lan W\ran/\tau$ as a function of the cycle period $\tau$. (b)Plot of $\lan\eta\ran$ and $\bar \eta$ as a function of the cycle period $\tau$. Here $\gamma_{fb}=0$}
\label{fig2}
\end{figure}

\begin{figure}[ht]
\centering
\includegraphics[width=1.1\columnwidth]{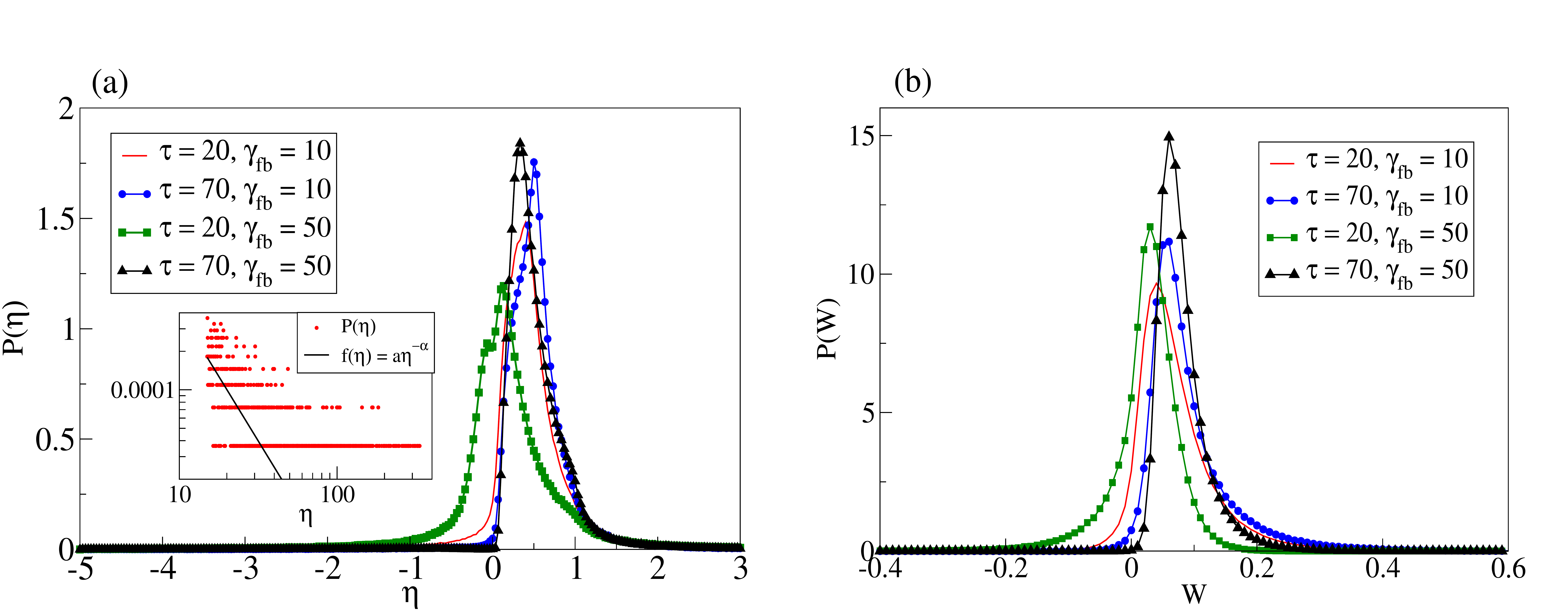}
\caption{(color online)(a)Plot of distribution  $P(\eta)$ vs $\eta$ for different values of $\gamma_{fb}$ and $\tau$. Distribution has power law tails with exponent close to $2$ (inset of (a)).(b) Plot of work distribution for different values of $\gamma_{fb}$ and $\tau$.}
\label{fig3}
\end{figure}
{\textcolor{black}{We are interested in stochastic quantities like total work done $W=W_1+W_2+W_3+W_4$ along a trajectory, where $W_i$'s are work performed in four steps of the protocol, and heats exchanged $Q_1$ and $Q_2$ during expansion and compression processes respectively. Internal energy of the Langevin system is given by $U(x,\dot{x})=\frac{1}{2}\ m\dot{x}^2 +\frac{1}{2}\ k(t) x^2$. Using Stochastic Thermodynamics \cite{Sekimoto97}, we can also find out expressions for work done and heat exchanged in all the steps of the protocol. In the first expansion process work done and heat exchanged are given by $W_1=\int_{0}^{\tau/2}\frac{1}{2}\ \dot{k}_1(t) x^2(t) ~dt$ and $Q_1=-\int_{0}^{\tau/2}(-\gamma \dot{x} +\sqrt{\gamma T}\xi(t)) \dot{x}(t)~dt$ respectively. In the instantaneous expansion step no heat exchange takes place and work done is nothing but change in the internal energy $W_2 = \frac{1}{2} (k_2(\tau/2)-k_1(\tau/2))x^2(\tau/2)$. In the third step again work and heat definitions remain as in the first step but with modified 
friction term and thus $W_3=\int_{\tau/2}^{\tau}\frac{1}{2}\ \dot{k}_2(t) x^2(t) ~dt$ and $Q_2=-\int_{\tau/2}^{\tau}(-(\gamma_{fb}+\gamma) \dot{x} +\sqrt{\gamma T}\xi(t)) \dot{x}~dt$. Fourth step, being instantaneous, again gives no heat exchange but work done is given by the change in the internal energy as, $W_4  = \frac{1}{2} (k_1(0)-k_2(\tau))x^2(\tau)$. We calculate the stochastic efficiency defined over a single trajectory of the Brownian particle (working medium of the micro machine) as, $\eta=W/(-Q_1)$. Two different averages of $\eta$ over the cycles can be calculated as,}} 

\beqa
\bar\eta = \frac{\lan W\ran}{-\lan Q_1\ran }\ ,~~~ \lan\eta\ran = \left\lan\frac{W}{-Q_1}\ \right\ran. 
\eeqa
{\textcolor{black}{As the system is subjected to large thermal fluctuations these two averages of stochastic efficiency are unequal except in large cycle time limit.}}

 
{\textcolor{black}{In simulations we integrate the Langevin equations Eq. \ref{Lang1} or Eq. \ref{Lang2}, depending on whether the trap expanding or contracting, by a velocity Verlet algorithm with Stratonovich discretization having time step $dt\sim10^{-3}$ and find average work and heat exchanged. These averages are over $10^4$ cycles of $k(t)$, after driving the system in the steady state. According to our convention, work done {\it by the system} and heat flow {\it into the bath} are taken to be positive.}}


{\textcolor{black}{We calculate the average work-output $\langle W\rangle$ from the system for varying feedback parameter $\gamma_{fb}$ and cycle time $\tau$ [Fig(1)]. With constant $\tau$, as we take $\gamma_{fb}$ beyond a certain value (see [Fig(1a)]), $\langle W\rangle$ decreases. For larger cycle time the decrease of $\langle W\rangle$ is slower than that of the smaller cycle time. Corresponding average stochastic efficiency $\langle \eta\rangle$ also decreases with $\gamma_{fb}$. Fig(1b) shows that $\langle W\rangle$ and $\langle \eta\rangle$ increases with cycle time $\tau$. In [Fig(2a)] average power, defined as $\langle W\rangle/\tau$, is plotted with $\tau$ for constant $\gamma_{fb}>0$ that shows non-monotonic behavior of power, as expected from engine. The power goes to zero in quasistatic limit as well as in high frequency i.e. small $\tau$ limit. Behavior of $\langle \eta\rangle$ and $\bar\eta$  with varying $\tau$ and $\gamma_{fb}>0$ are shown in [Fig(2b)]. It shows that both $\langle \eta\rangle$ and $\bar\eta$ initially increases with $\tau$ and finally saturate for larger $\tau$. Though the values at which they saturate for large $\tau$ are different and $\langle \eta\rangle$ overshoots $\bar\eta$. In [Fig(3a)] and [Fig(3b)], distributions $P(\eta)$ and $P(W)$ are shown for different combinations of $\gamma_{fb}$ and $\tau$. In the inset we have shown that the tail of $P(\eta)$ goes as $\eta^{-\alpha}$ with $\alpha\simeq 2$. The stochastic efficiency is unbounded and distribution is very broad and it shows power law tail with exponent around 2. It does not contain any prominent local minima at any particular efficiency. Fluctuations in $\eta$ are large \cite{Arun14,ArunPhysicaA,ArunIJMPB}. In fact relative variance of the stochastic efficiency is much larger than mean value. This implies that the average quantity is not a good physical variable here. In such situations one has to study the full probability distribution of $\eta$. 
}  



{\textcolor{black}{We have considered a single Brownian particle kept in a time-dependent harmonic trap as discussed in reference \cite{Arun14}. Though in the model discussed here, instead of two heat baths we have a single bath at temperature $T$. However we apply a velocity feed back protocol during compression step of the trap, where the friction term is replaced by $-( \gamma_{fb}+\gamma)v$. Here $\gamma_{fb}>0$ is the constant drag that acts on the particle due to the feedback. Therefore, the effective friction coefficient during the compression is large compared to that of in the expansion process. Though the temperature of the bath is constant through out the dynamics, due to the effective, large friction from the feedback, the heat loss during compression is more than the expansion process. So, to the particle, the effective temperature $T_{eff}$ of the surroundings appears to be smaller than $T$. Hence we can extract work from single heat bath without violating the  Second Law of thermodynamics. This technologically is extremely important which allows extracting work from a single bath with instantaneous (i.e. without any delay) velocity dependent feedback. We note here that the power-law exponent $\alpha\simeq 2$ obtained from the tail of $P(\eta)$ here, is also obtained in various other micro machines as, (i) in \cite{VerleyPRL, Arun14, ArunPhysicaA, ArunIJMPB} (ii) in case of a spin-1/2 system coupled to two heat baths simultaneously \cite{basu} and (iii) in case of a micro-heat engine with a Brownian particle driven by micro adiabatic protocol\cite{Edgar16,saha}. Our feed backed engine performance is dominated by fluctuations and hence it is not a reliable engine. Presently we are studying feedback controlled engines as in the present case, however with an optimal protocol. This may increase the performance characteristics of the engine.}}

\section{Acknowledgement} 


AS thanks UGCFRP and RMS thanks DST, India for financial support. AMJ also thanks DST, India for J. C. Bose National Fellowship. AS thanks Edgar Roldan for initial discussions on micro heat engines. Authors thank P.S.Pal for computation. 


\end{document}